\documentclass[11pt]{article}
\usepackage[dvipsnames,x11names]{xcolor} 
\usepackage{lscape}
\usepackage{amssymb} 
\usepackage{graphicx}
\usepackage{footnote}
\usepackage[utf8]{inputenc}	
\usepackage{mathtools}
\usepackage{amsthm}
\usepackage{amsfonts}
\usepackage[english]{babel}
\usepackage{bigints}
\usepackage{mathrsfs}
\usepackage{lineno}
\usepackage{nicefrac}
\usepackage{enumerate}
\usepackage{bbm}

\usepackage{listings}

\usepackage{commath}
\usepackage{multirow}

\usepackage{newtxtext}
\usepackage[T1]{fontenc}

\usepackage{algorithm} 
\usepackage{algpseudocode} 
\usepackage{booktabs}

\usepackage[most]{tcolorbox}
\usepackage{mdframed}

\usepackage[]{natbib}
\usepackage{csquotes}

\usepackage{tabularx}

\usepackage[colorlinks=true,citecolor=blue,linkcolor=black,urlcolor=blue]{hyperref}
\definecolor{light-gray}{gray}{0.91}

\usepackage[margin=2cm]{geometry}

\usepackage{sectsty}
\sectionfont{\color{BrickRed}}

\sectionfont{\color{RoyalBlue4}}
\subsectionfont{\color{RoyalBlue4}}
\subsubsectionfont{\color{RoyalBlue4}}
\paragraphfont{\color{RoyalBlue4}}

\allowdisplaybreaks[1]
\usepackage{pifont}

\makeindex

\newcommand{\imag}{\textbf{i}}

\newcommand{\Lagr}{\mathcal{L}}

\begin{document}

\begin{flushleft}

{\huge\bfseries \color{DodgerBlue4}{Higher--order gradient modeling of velocity dispersion in tight sandstones with implications to agroseismology}\par} 
	\vspace{1.5em}  

	Rafael Abreu\textsuperscript{1,*}, Ashay Motghare\textsuperscript{2}, Chahana Nagesh\textsuperscript{1}, Alireza Nadi\textsuperscript{3} and Hassam Khan\textsuperscript{4}  \\[0.5em]  
	
	\scriptsize 
	\textsuperscript{1} \textit{Institut de Physique du Globe de Paris, CNRS, Universit\'e de Paris, Paris, France} \\ 
	\textsuperscript{2} \textit{Indian Institute of Science Education and Research Kolkata, West Bengal, India} \\
	\textsuperscript{3} \textit{Department of Mining Engineering, Faculty of Engineering, Tarbiat Modares University, Tehran, Iran} \\ 
	\textsuperscript{4}  \textit{Department of Mathematics and Statistics, Institute of Business Management, Sindh, Pakistan} \\ [1em]   
	\textsuperscript{*} {email: rabreu@ipgp.fr}
	
	\normalsize 
\end{flushleft}

\begin{abstract}
	Modeling frequency-dependent shear-wave propagation in fluid-saturated porous rocks remains a challenge today because existing poroelastic theories generally rely a large number of constitutive parameters that are difficult to constrain from seismic observations. We propose a dynamically consistent higher-gradient continuum model that reproduces shear-wave dispersion using only three effective material parameters: the macroscopic shear velocity and two characteristic length scales governing higher-order elasticity and gradient inertia. We validate the model against laboratory measurements of glycerin-saturated tight sandstone over a wide range of effective pressures. The inversion reproduces the measured shear-wave dispersion, including positive, weak and negative dispersion regimes. We finally discuss the relationship between the proposed formulation and micropolar theories, the computational challenges and the potential of reduced-order generalized continuum models for seismic exploration, reservoir characterization, and emerging agroseismology applications.
\end{abstract}


\newpage

\section{Introduction}

Seismic modeling of wave propagation including the complex nature of micro-structure remains a challenge today and there are several reasons for this. On the one hand, taking into account the presence of micro-structure in the equations of motion requires the introduction of new variables, which unavoidably bring with them new physical parameters that have to be determined. This means that properly accounting for micro-structural effects in seismic wave propagation carries not only theoretical challenges but experimental ones as well.

In principle, one can include as many parameters as desired in a theoretical description and estimate them from physical or experimental observations. It may therefore seem that theoretical limitations can simply be overcome by performing sufficiently detailed experiments and properly tuning the corresponding parameters. However, the inverse problem, namely obtaining physically meaningful parameters from experimental observations, becomes increasingly difficult as new variables are introduced into the theoretical description, and in many cases it becomes an ill-posed problem.

The determination of too many elastic parameters from a limited set of experimental observations can easily become ill-posed. For this reason, it is important to look for realistic descriptions of seismic wave propagation that reproduce the correct physics with the smallest possible number of parameters. In the particular case of seismic wave propagation including poroelastic effects, where the pore space is filled with fluids, the standard theory is that of Biot \citep{Biot1955,Biot1956,Biot1956b}. However, when modeling shear-wave dispersion, the original Biot theory can become insufficient. For this reason, several modifications have been proposed, some of which require up to seventeen material parameters \citep[e.g.][]{kimura2006frame,kimura2008experimental,kimura2013shear}.

It thus becomes evident that new theoretical models of seismic wave propagation, capable of reproducing the observed physics with a reduced number of parameters, are needed. In this work, we show that the a higher-order gradient model is able to reproduce the shear-wave dispersion experimentally observed in a tight sandstone using a substantially reduced number of parameters compared to the modified Biot model. Additionally to reproducing the experimental observations, we also provide a physical interpretation of the higher-order gradient parameters, allowing them to be related directly to observable features of the dispersion curves. This interpretation establishes a direct connection between the mathematical formulation of the model and the underlying wave physics. We first present the theory with paying particular attention to the physical interpretation of its parameters. We then validate the model against experimental observations in tight sandstones. Finally, we discuss the obtained results and their implications for the emerging field of agroseismology.

\section{Higher-order gradient elasticity}
\label{sec.micro_to_macro}

The present formulation belongs to the class of dynamically consistent gradient-elastic models \citep[e.g.][]{metrikine2002one,Askes2002,Askes2007}, in which higher-order elastic stiffness and higher-order inertia are introduced simultaneously through two independent characteristic lengths. The strain tensors $\varepsilon,\eta$ are given by
\begin{align}
	\varepsilon_{ij}
	&=
	\frac{1}{2}
	\left(
	\partial_j u_i
	+
	\partial_i u_j
	\right),
	\quad 
	\eta_{ijk}=
	\partial_k \varepsilon_{ij}
	=
	\frac{1}{2}
	\left(
	\partial_k\partial_j u_i
	+
	\partial_k\partial_i u_j
	\right),
	\label{eq.Gradient_Strain_Measures}
\end{align}
where $u:\mathbb{R}^3 \rightarrow \mathbb{R}^3$ is the displacement vector. Assuming no coupling between the classical strain and the strain gradient, the elastic-energy density is given by
\begin{align}
	\mathcal{E}
	&=
	\frac{1}{2}
	\mathbb{C}_{ij\ell m}
	\varepsilon_{ij}\varepsilon_{\ell m}
	+
	\frac{1}{2}
	(L_c^2)_{kn}
	\mathbb{C}_{ij\ell m}
	\eta_{ijk}\eta_{\ell mn}.
	\label{eq.Gradient_Elastic_Energy_General}
\end{align}
where $\mathbb{C}$ is a fourth order tensor of elastic constants and $L_c^2$ is a symmetric second-order constitutive tensor with dimensions of length squared. The total kinetic energy density is defined as follows
\begin{align}
	K
	&=
	\frac{\rho}{2}
	\partial_t u_i\,\partial_t u_i
	+
	\frac{\rho L_t^2}{2}
	\partial_j\partial_t u_i\,
	\partial_j\partial_t u_i.
		\label{eq.Gradient_strain_Kinetic_Energy}
\end{align}
where the $\rho$ is the mass density and $L_t$ is the characteristic inertial length. Using eqs. \eqref{eq.Gradient_Elastic_Energy_General} and \eqref{eq.Gradient_strain_Kinetic_Energy} and solving the minimization problem for the Langrangian formulation which defines the action integral $I$ of the form
\begin{flalign}
	I = \int_0^t \int _{\Omega} \Lagr \dif x\dif t = \int_0^t \int _{\Omega} (K - \mathcal{E}) \dif x \dif t 
	,
	\label{eq.Minimization_Functional}
\end{flalign}
where $\Omega$ denotes the volume of the continuum,  $\Lagr$ is the Lagrangian defined as the difference between the kinetic energy $K$ and the potential energy $\mathcal{E}$. We can find the equations of motion given by the following expressions
\begin{align}
	\rho
	\left(
	\delta_{ij}
	-
	L_t^2\delta_{ij}\partial_k\partial_k
	\right)
	\partial_t^2u_j
	&=
	\partial_j\sigma_{ij}
	-
	\partial_j\partial_k\tau_{ijk} ,
	\label{eq.Gradient_Equation_Motion_General}
\end{align}
where the stress tensors are given by the following expressions
\begin{align}
	\sigma_{ij}
	=
	\frac{\partial\mathcal{E}}
	{\partial\varepsilon_{ij}}
	=
	\mathbb{C}_{ij\ell m}\varepsilon_{\ell m},
	\qquad
	\tau_{ijk}
	=
	\frac{\partial\mathcal{E}}
	{\partial\eta_{ijk}}
	=
	(L_c^2)_{kn}
	\mathbb{C}_{ij\ell m}
	\eta_{\ell mn}.
	\label{eq.Gradient_Stress_Tensors}
\end{align}
The free-surface boundary conditions associated are
\begin{align}
	\underbrace{
		n_j \sigma_{ij}
		-
		n_j \partial_k \tau_{ijk}
	}_{\text{elastic traction}}
	+
	\underbrace{
		\rho L_t^2 n_j \partial_j \partial_t^2 u_i
	}_{\text{gradient-inertia traction}}
	= 0,
	\qquad 
	\underbrace{
		n_k \tau_{ijk}
	}_{\text{higher-order traction}}
	= 0.
\end{align}
The first equation represents the generalized traction condition, including both the elastic and gradient-inertia contributions, while the second corresponds to the vanishing higher-order (double-force) traction. In 3D isotropic media we can write 
\begin{align}
	\sigma_{ij}
	=
	\lambda\delta_{ij}\partial_\ell u_\ell
	+
	\mu
	\left(
	\partial_i u_j
	+
	\partial_j u_i
	\right),
	\qquad 
	\tau_{ijk}
	=
 L_c^2 \partial_k\sigma_{ij} ,
	\label{eq.Isotropic_Gradient_Stresses}
\end{align}
where $\lambda,\mu$ are Lam\'e parameters. In a 1D shear isotropic medium we can write the following wave equation
\begin{align}
	\rho
	\left(
	\partial_t^2u
	-
	L_t^2\partial_t^2\partial_x^2u
	\right)
	&=
	\mu\partial_x^2u
	-
 \mu L_c^2 
	\partial_x^4u.
	\label{eq.Gradient_Wave_Equation_1D}
\end{align}

\subsection{Dispersion analysis}

If we assume plane wave propagation of the form
\begin{align}
	u=  A_1 e^{\imag(k x + \omega t)},
	\label{eq.Plane_Waves1}
\end{align}
where $ A_1 $ is the (unknown) amplitude, $k$ the wavenumber and $\omega$ the angular frequency. We can write eq. \eqref{eq.Gradient_Wave_Equation_1D} as follows  
\begin{align}
	\mathcal{A} \cdot \overline{\gamma} = 0 .
	\label{eq.Dispersion_Determiant_Relation}
\end{align}

In order to obtain non-trivial solutions $(\mathcal{A} \neq 0)$ of eq. \eqref{eq.Dispersion_Determiant_Relation}, one must impose that
\begin{align}
	\text{det} \, \overline{\gamma}  = 0.
	\label{eq.Det_Algebraic_System}
\end{align}
Equation \eqref{eq.Det_Algebraic_System} is called the dispersion relation from which we can obtain an expression for the general phase velocity $(\omega/k)$ given by the following expression
\begin{align}
	v_s^2	&=
	\frac{
		c_0^2-L_t^2\omega^2
		+
		\sqrt{
			\left(
			c_0^2-L_t^2\omega^2
			\right)^2
			+
			4c_0^2L_c^2\omega^2
		}
	}{2}, \qquad c_0^2 = \frac{\mu}{\rho} ,
	\label{eq.Gradient_Phase_Velocity_Frequency}
\end{align}
with the two limiting cases
\begin{align}
	\lim_{\omega\to 0}v_{s}^2=c_0^2, \qquad \lim_{\omega\to \infty}v_{s}^2=c_0^2 \frac{L_c^2}{L_t^2}.
	\label{eq.general_micro_macro}
\end{align}

We can interpret micro and macro scales as the two limiting (frequency) cases of the phase velocity eq. \eqref{eq.general_micro_macro} as follows 
\begin{align}
	\begin{aligned}
		\text{macroscale} \to \lim_{\omega\to0}v_{s}^2 &= \frac{\mu}{\rho}=v_{\text{macro}}^2,\\
		\text{microscale} \to \lim_{\omega\to\infty}v_{s}^2 &= \frac{\mu}{\rho} \frac{L_c^2}{L_t^2}=v_{\text{micro}}^2.
	\end{aligned}
	\label{eq.Phase_Velocity_limit}
\end{align}

We can next write a relationship between macroscopic and microscopic phase velocities, by combining eqs. \eqref{eq.Phase_Velocity_limit}, which results in the following expression
\begin{align}
	v_{\text{micro}}^2 = v_{\text{macro}}^2 \frac{L_c^2}{L_t^2} .
	\label{eq.scaling}
\end{align}

Equation \eqref{eq.scaling} provides a mechanical relationship between micro and macro scales. We can write eq. \eqref{eq.Gradient_Phase_Velocity_Frequency} as follows
\begin{align}
	v_s^2
	&=
	\frac{
		v_{\text{macro}}^2
		-
		L_t^2\omega^2
		+
		\sqrt{
			\left(
			v_{\text{macro}}^2
			-
			L_t^2\omega^2
			\right)^2
			+
			4v_{\text{macro}}^2L_c^2\omega^2
		}
	}{2}.
	\label{eq.Gradient_Phase_Velocity_Macro}
\end{align}
If we call
\begin{align}
	\Omega
	=
	\frac{\omega \,  L_t}{v_{\text{macro}}},
	\qquad
	R
	=
	\frac{v_{\text{micro}}}
	{v_{\text{macro}}} = \frac{L_c}{L_t},
\end{align}
the (normalized) phase velocity (eq. \eqref{eq.Gradient_Phase_Velocity_Macro}) becomes
\begin{align}
	\frac{v_s^2}{v_{\text{macro}}^2}
	&=
	\frac{
		1-\Omega^2
		+
		\sqrt{
			\left(1-\Omega^2\right)^2
			+
			4R^2\Omega^2
		}
	}{2}.
	\label{eq.Gradient_Phase_Velocity_Normalized}
\end{align}

\begin{figure}
	\begin{center}
		\includegraphics[width=1\textwidth]{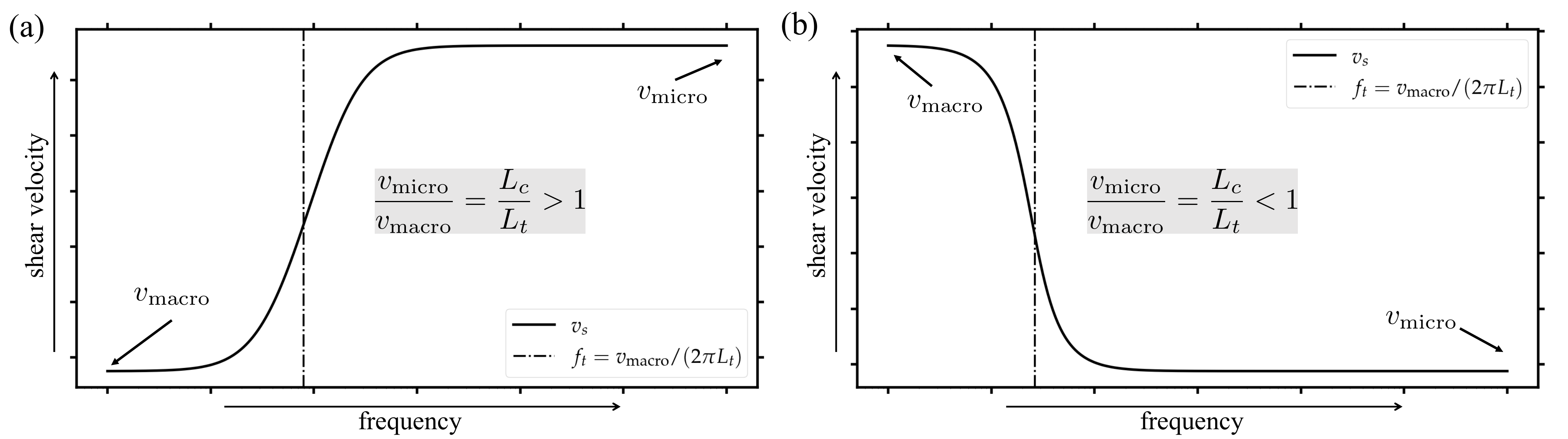}
		\caption{Shear velocity dispersion as a function of frequency for: (a) $\nicefrac{v_{\text{micro}}} {v_{\text{macro}}}>1$ and (b) $\nicefrac{v_{\text{micro}}} {v_{\text{macro}}}<1$.}
		\label{fig.micro_macro_excess}
	\end{center}
\end{figure}

Note that $R$ controls the magnitude and direction of the shear-wave dispersion. When $R>1$, the phase velocity increases from the macroscopic velocity at low frequencies to a larger microscopic velocity at high frequencies (see Fig. \ref{fig.micro_macro_excess}--a). When $R=1$, the two limiting velocities coincide and the medium is non-dispersive. When $0<R<1$, the phase velocity decreases from the macroscopic velocity at low frequencies to a smaller microscopic velocity at high frequencies (see Fig. \ref{fig.micro_macro_excess}--b).

\subsection{Physical interpretation of higher-order gradient parameters}

The presented strain-gradient model contains two characteristic lengths: the elastic characteristic length $L_c$ and the inertial characteristic length $L_t$.

\paragraph{The inertial characteristic length $L_t$}

The characteristic length $L_t$ enters the kinetic-energy in the velocity-gradient contribution (see eq. \eqref{eq.Gradient_strain_Kinetic_Energy}). This means that $L_t$ controls the influence of gradient inertia relative to classical translational inertia. At the characteristic transition frequency defined as
\begin{align}
	\omega_t
	&=
	\frac{v_{\mathrm{macro}}}{L_t}
	=
	\frac{1}{L_t}
	\sqrt{\frac{\mu}{\rho}}.
	\label{eq.Gradient_Transition_Frequency}
\end{align}
The quantity $\omega_t$ represents the frequency scale at which the classical and gradient-inertia contributions become comparable.

\paragraph{The elastic characteristic length $L_c$}

The characteristic length $L_c$ enters the elastic-energy density through the strain-gradient contribution, through the coefficient of the fourth-order spatial derivative. This means that $L_c$ controls the magnitude of the higher-gradient elastic stiffness. Increasing $L_c$ increases the resistance of the material to spatial variations of strain and strengthens the fourth-order elastic contribution.

Simply, the characteristic lenght $L_c$ characterizes the spatial scale below which strain-gradient effects become significant. For deformation varying over distances much larger than $L_c$, the higher-gradient elastic contribution is small and the classical elastic response is recovered.

\paragraph{Relationship between the two characteristic lengths}

The low- and high-frequency limiting shear velocities $v_{\mathrm{micro}}$ and $v_{\mathrm{macro}}$ are controlled by the ratio $L_c/L_t$ (see eq. \eqref{eq.scaling}). As a consequence the ratio $L_c/L_t$ controls the magnitude (and direction) of the velocity dispersion. We can observe that,
\begin{align}
\begin{aligned}
		L_c>L_t
	&\quad\Longrightarrow\quad
	v_{\mathrm{micro}}>v_{\mathrm{macro}},
	\\
	L_c=L_t
	&\quad\Longrightarrow\quad
	v_{\mathrm{micro}}=v_{\mathrm{macro}},
	\\
	L_c<L_t
	&\quad\Longrightarrow\quad
	v_{\mathrm{micro}}<v_{\mathrm{macro}}.
	\label{eq.Micro_Macro_High_Gradient_relations}
\end{aligned}
\end{align}

Therefore, $\omega_t$ is the characteristic frequency scale of the transition (see eq. \eqref{eq.Gradient_Transition_Frequency}), but the ratio $L_c/L_t$ determines the difference between the low- and high-frequency limiting velocities.
 
\section{Experimental validations in tight sandstones}

Tight sandstone reservoirs are sedimentary formations that have the potential of storing hydrocarbons \citep{Zou2010,ZOU2015,JIA2012139,zoback2019unconventional,li2015review}. In contrast to conventional oil reservoirs, tight sandstones reservoirs have experienced prolonged mechanical compaction and cementation, which reduce pore space and increase rock stiffness as mineral precipitates progressively fill inter-granular voids \citep{morad2010impact,wilson2014influence}. As a consequence, these unconventional characteristics introduce new challenges for seismic exploration and reservoir imaging, since tight formations require more sophisticated theoretical descriptions due to their complex pore geometry and weak seismic responses \citep{zou2012tight,Mavko2009,Avseth2010}. 

Laboratory measurements on saturated tight sandstones therefore provide an ideal benchmark for testing theoretical models of wave propagation. They establish the experimental link between pore-scale fluid flow mechanisms and the corresponding seismic observables, enabling quantitative interpretation of geophysical measurements and improved reservoir characterization and imaging \cite{yin2017pressure,ba2023effect}, a tight sandstone sample characterized by a crack--pore microstructure has been investigated under different saturation conditions, including dry (nitrogen), brine, and glycerin. Experiments by \cite{yin2017pressure}  were designed to investigate the frequency dependence of the elastic properties of fully saturated tight sandstones over a broad frequency range of 2–200 Hz and ultrasonic 1 MHz frequencies.

The experiments were performed on a cylindrical tight sandstone core from the Xujiahe Formation (China), characterized by approximately $9\%$ porosity, low permeability, and a significant crack--pore microstructure. Such rocks are representative of unconventional hydrocarbon reservoirs where wave-induced local fluid flow between compliant microcracks and stiff pores produces significant elastic dispersion \cite{yin2017pressure}.

At seismic frequencies (2--200 Hz), the elastic response was measured using a forced-oscillation apparatus \cite{yin2017pressure}. These measurements provided frequency-dependent Young's modulus, Poisson's ratio and attenuation. The same specimen was subsequently tested using ultrasonic transducers operating at approximately 1 MHz to determine the high-frequency elastic moduli from compressional and shear-wave travel times. Measurements were repeated under several effective pressures for dry (nitrogen), brine, and glycerin saturation.

It is important to emphasize that existing poroelastic theories \citep[e.g.][]{ba2023effect,kimura2006frame,kimura2008experimental,kimura2013shear} generally rely on detailed representations of pore geometry and fluid flow, requiring a relatively large number of parameters (up to seventeen in some cases). These models provide an accurate description of laboratory experiments, however, many of these parameters are difficult or impossible to constrain from field-scale seismic observations. This is the main motivation the for the reduced-order higher-gradient description that, as we will next see, reproduces the essential dispersive shear behavior while involving only a reduced number of material parameters.

From all measured samples by \cite{yin2017pressure}, we select the glycerin-saturated case corresponding to (i) 40$^{\circ}$C and at the pressures of 2 MPa, 5 MPa, 7 MPa, 10 MPa, 15 MPa, 20 MPa, 25 MPa, and 30 MPa. These cases are particularly interesting because the theoretical squirt-flow model presented in \cite{yin2017pressure} does not accurately reproduce the measured shear-wave dispersion, thereby providing a stringent benchmark for evaluating the proposed higher-gradient model. 

\begin{figure}
	\begin{center}
		\includegraphics[width=1\textwidth]{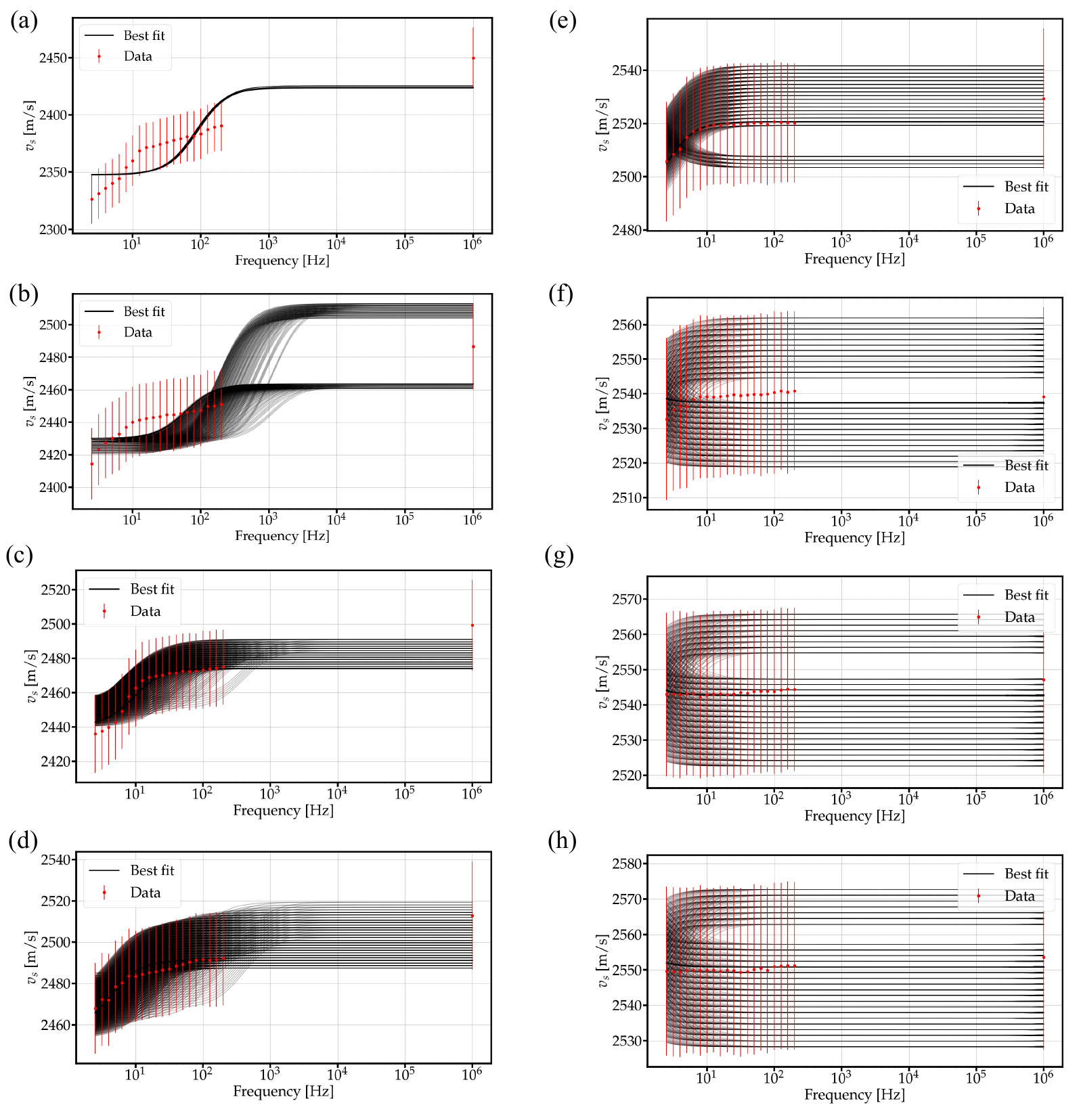}
		\caption{Frequency-dependent shear-wave velocity data (from \cite{yin2017pressure}) with error bars for glycerin-saturated at  $40^\circ$C under different pressures: (a) 2 MPa, (b) 5 MPa, (c) 7 MPa, (d) 10 MPa, (e) 15 MPa, (f) 20 MPa, (g) 25 MPa, and (h) 30 MPa.}
		\label{fig.experimentsALL}
	\end{center}
\end{figure}

The corresponding inversion results are shown in Figs.~\ref{fig.experimentsALL} and Table \ref{tab:velocity_length}. 
The inversion results show that the higher-gradient model reproduces the measured shear-wave velocities over the complete investigated pressure range. The agreement is particularly good considering that the model contains only three independent fitting parameters: the macroscopic limiting velocity, the microscopic-to-macroscopic velocity ratio, and one characteristic length. The small values of $\chi^2$ indicate that the predicted phase velocities remain close to the experimental measurements within the reported uncertainties (see Table \ref{tab:velocity_length}).
\begin{table}
	\centering
	\caption{Best-fitting microscopic and macroscopic phase velocities together with the corresponding characteristic lengths obtained from the higher-gradient model. The corresponding transition frequency is defined as $f_t=v_{\rm macro}/(2\pi L_t)$.}
	\label{tab:velocity_length}
	\begin{tabular}{ccccccc}
		\toprule
		Pressure [MPa] &
		$v_{\rm micro}$ [m/s] &
		$v_{\rm macro}$ [m/s] &
		$L_t$ [m] &
		$L_c$ [m] &
		$f_t$ [Hz] &
		$\chi^2$ \\
		\midrule
		2  & 2423.91 & 2347.74 & 4.4542  & 4.5987  & 83.95 & 10.60 \\
		5  & 2463.52 & 2430.00 & 6.4216  & 6.5102  & 60.25 & 3.24 \\
		7  & 2474.07 & 2440.41 & 42.0290 & 42.6087 & 9.24  & 1.35 \\
		10 & 2491.13 & 2457.24 & 90.6678 & 91.9184 & 4.32  & 0.88 \\
		15 & 2520.67 & 2486.38 & 171.32  & 173.68  & 2.31  & 0.12 \\
		20 & 2537.38 & 2555.00 & 604.16  & 600.00  & 0.67  & 0.26 \\
		25 & 2542.69 & 2560.34 & 551.16  & 547.36  & 0.74  & 0.05 \\
		30 & 2550.80 & 2568.52 & 604.16  & 600.00  & 0.67  & 0.64 \\
		\bottomrule
	\end{tabular}
\end{table}

The degree to which the individual parameters are constrained, however, varies strongly depending on the effective pressure. At low pressures, there is a clear dispersive transition between microscopic and macroscopic velocities. This allows both the velocity contrast and the characteristic frequency scale to be determined. As the pressure increases, the measured dispersion, i.e.e, the difference between micro and macro velocities, becomes weaker and the inferred transition frequency moves below the seismic-frequency interval. This produces that the observations constrain primarily the limiting velocities and the ratio $L_c/L_t$, leaving the absolute values of $L_c$ and $L_t$ poorly constrained/resolved.

\section{Discussion}

\subsection{Experimental results}

Figure \ref{fig.experimentsALL} shows that at low effective pressures the laboratory measurements exhibit a clear increase of shear-wave velocity with frequency, i.e., $v_{\rm micro}>v_{\rm macro}$. For instance, at $2$ MPa the velocity increases from $v_{\rm macro}=2348~\mathrm{m/s}$ to $v_{\rm micro}=2424~\mathrm{m/s}$. This difference progressively decreases with increasing pressure. This behavior can be attributed to wave-induced local fluid flow between cracks and stiffer pores. At low frequencies, the pore fluid has sufficient time to redistribute during one loading cycle.  As the frequency increases, the fluid has progressively less time to redistribute between compliant cracks and stiffer pores, producing an increase in the apparent stiffness of the saturated rock and consequently the shear-wave velocity. 

Note that the inferred microscopic and macroscopic velocities progressively converge with increasing effective pressure, this indicates that the dispersion weakens as the rock stiffens. For pressures up to approximately $15$ MPa, the microscopic velocity can be larger than the macroscopic velocity $(v_{\rm micro} >v_{\rm macro})$, and the phase velocity therefore increases with frequency. At pressures of $20$ MPa and above, the inversion shows that $v_{\rm micro} < v_{\rm macro}$, which correspond to a weak decrease of phase velocity with frequency. Such behavior is naturally accommodated by the dynamically consistent higher-gradient formulation because the elastic and inertial length scales are independent.

The inversion also reveals a clear evolution of the inferred characteristic lengths with increasing effective pressure. Unlike the limiting velocities, which remain well constrained by the experimental observations, the inversion predicts a significant increase of the characteristic lengths as the dispersion becomes weaker. We emphasize that this behavior should not be interpreted as a direct increase of the physical dimensions of pores or cracks. Instead, $L_c$ and $L_t$ are effective continuum parameters that determine the characteristic frequency over which the transition between the macroscopic and microscopic velocities occurs. Instead, $L_c$ and $L_t$ are effective continuum parameters that determine the characteristic frequency over which the transition between the macroscopic and microscopic velocities occurs. When the measured phase-velocity curve becomes nearly frequency independent, the inversion can no longer uniquely constrain the absolute values of these characteristic lengths.

This interpretation becomes clearer by introducing the characteristic transition frequency
\begin{align}
	f_t=\frac{v_{\rm macro}}{2\pi L_t},
\end{align}
which controls the frequency at which the transition between the macroscopic and microscopic limiting velocities occurs. At low effective pressures, the inferred values of $f_t$ lie within the experimental frequency band, allowing the complete dispersive transition to be observed. 

At higher effective pressures, the measured phase-velocity curves become nearly frequency independent. In this regime, the material response approaches classical linear elasticity, with
\begin{align}
	v_{\rm micro}
	\approx
	v_{\rm macro}.
\end{align}
Accordingly, the ratio of the characteristic lengths approaches unity,
\begin{align}
	\frac{L_c}{L_t}
	\approx
	1,
\end{align}
and the higher-order elastic and inertial contributions become balanced in the dispersion relation. Because the phase-velocity curve is nearly flat, the experimental data no longer contain sufficient information to determine the absolute values of $L_c$ and $L_t$ independently. The large inferred values of the two lengths at high pressure should therefore not be assigned a direct physical meaning; only their ratio, which controls the difference between the microscopic and macroscopic limiting velocities, remains constrained. Thus, the high-pressure results should be interpreted primarily as the recovery of an effectively classical elastic response rather than as evidence for a particular transition frequency or characteristic length scale.

Overall, the results demonstrate that the higher-order gradient model reproduces three distinct regimes using the same governing equation: pronounced positive dispersion at low pressure, weak dispersion at intermediate pressure, and slightly negative dispersion at high pressure. The proposed higher-gradient formulation therefore provides a compact continuum description of frequency- and pressure-dependent shear-wave dispersion while requiring only three effective material parameters. Despite its simplicity, the model accurately reproduces the laboratory observations over the entire investigated pressure range without explicitly modeling pore-scale fluid flow.

\subsection{S--wave dispersion: limitations of classical Biot's models}

It is well known that Biot's theory allows to accurately model P-wave dispersion but encounters important challenges for modeling S-wave dispersion  \citep[e.g.][]{ba2023effect}. This limitation is related to fluid–solid coupling assumed in Biot's theory, for which S-waves are weakly sensitivity. In this sense, the limitations of Biot's theory for modeling shear wave dispersion is mathematically structural as we will next see.

The 1D Biot's shear equations of motion can be written as follows \citep{bourbie1987acoustics,van2013multi,Carcione2015}
\begin{align}
\begin{aligned}
		\left[(1-\phi)\rho_s+\phi\rho_f\right]\partial_t^2 u-\rho_f\partial_t^2 w & =\mu\partial_x^2 u,
	\\
	\rho_f\partial_t^2 u-\frac{\alpha_\infty\rho_f}{\phi}\partial_t^2 w-\frac{\eta F}{\kappa}\partial_t w& =0,
\end{aligned}
	\label{eq.Biot_Equation_Explicit}
\end{align}
where $u$ is the transverse displacement of the solid skeleton and $w$ is the relative displacement of the pore fluid with respect to the solid. The parameter $\phi$ is the porosity, $\rho_s$ is the grain density, $\rho_f$ is the pore-fluid density, $\alpha_\infty$ is the tortuosity, $\eta$ is the dynamic viscosity of the pore fluid, $\kappa$ is the permeability, and $F$ is the viscous correction factor. 

The bulk density of the saturated sediment $\rho$ can be written as follows
\begin{align}
	\rho=(1-\phi)\rho_s+\phi\rho_f,
\end{align}
and the dynamic mass coefficient associated with the relative fluid motion is
\begin{align}
	m=\frac{\alpha_\infty\rho_f}{\phi}.
\end{align}

Therefore, eq. \eqref{eq.Biot_Equation_Explicit} can be written in the compact form as follows
\begin{align}
\begin{aligned}	
	\rho\partial_t^2 u-\rho_f\partial_t^2 w & =\mu\partial_x^2 u,
\\
\rho_f\partial_t^2 u-m\partial_t^2 w-\frac{\eta F}{\kappa}\partial_t w & =0.
\end{aligned}
	\label{eq.Biot_Equation_Compact}
\end{align}
In the inviscid limit, the viscous drag term vanishes, i.e.,
\begin{align}
	\frac{\eta F}{\kappa}\partial_t w=0,
\end{align}
and eq. \eqref{eq.Biot_Equation_Compact} reduces to
\begin{align}
\begin{aligned}
		\rho\partial_t^2 u-\rho_f\partial_t^2 w & =\mu\partial_x^2 u \\
	\rho_f\partial_t^2 u-m\partial_t^2 w & =0.
\end{aligned}
	\label{eq.Biot_Equation_Inviscid}
\end{align}

This shows that, when viscosity is neglected, permeability no longer appears explicitly in the equations of motion. Eq. \eqref{eq.Biot_Equation_Inviscid} can be reduced to a general type equation given by
\begin{align}
	\left[ \partial_t ^2  - c_{x2}^2 \partial_x ^2 \right]  (u,w) = 0 ,
	\label{eq.General_EQ}
\end{align}
where the parameters are given by
\begin{align}
	c_{x2}^2=\frac{\mu\alpha_\infty}{\alpha_\infty\left[(1-\phi)\rho_s+\phi\rho_f\right]-\phi\rho_f}.
\end{align}
The general phase velocity is given by
\begin{align}
	v_s^2=\frac{\omega^2}{k^2}=  c_{x2}^2  .
	\label{eq.General_Phase_vel}
\end{align}
showing that, in the inviscid limit, the Biot's model cannot reproduce the frequency-dependent shear-wave velocities observed in the laboratory.

\subsection{Equivalence to micropolar models}

The micropolar model describes a continuum with additional rotational degrees of freedom $(\theta)$ which are different from the rotation computed using the curl of the displacement $(\nabla \times u)$ \citep{Eringen1999}. The model has been mainly used in seismology for the kinematic description of seismic sources \citep{Twiss01052009,JGRB:JGRB8837,teisseyre2006earthquake,Unruh1996,Unruh2003,Lewis2008,TWISS1991215,lewis2015seismogenic}. We will next consider the case for zero micro-inertia $I=0$; the micropolar model can be written as follows \citep{abreu2025micropolar}
\begin{align}
	\begin{aligned}
		\rho \, \partial_t ^2 u &  = (\mu + \mu_c) \partial_x ^2 u - 2 \mu_c \partial_x \theta , \\
		0 &  = \mu L_c^2 \partial_x ^2 \theta +  2\mu_c \partial_x u - 4\mu_c\theta ,
	\end{aligned}
	\label{eq.Izero_Equation_Motion_1D}
\end{align}
where $u:\mathbb{R}^3 \rightarrow \mathbb{R}^3$ is the transverse displacement vector,  $\theta:\mathbb{R}^3 \rightarrow \mathbb{R}^3$ the local/independent rotation, $\rho$ the material density, $I$ the micro-inertia density, $\mu$ the shear modulus, $\mu_c$ the Cosserat couple modulus \citep{neff2009new,neff2010} responsible for the resistance of the medium to the rotation of a particle, $L_c$ is a characteristic length. The elastic term $\mu L_c^2$ is a simplification of the original elastic constants required in the micropolar model \cite{Eringen1999} and it is equivalent to the hypothesis that $\nabla \cdot \theta$ and $\nabla \times \theta$ do not cause any stress in the medium \citep{neff2009new,neff2010,neff2010linear}. 

Eq. \eqref{eq.Izero_Equation_Motion_1D} can be interpreted as a quasi-static micropolar continuum. The displacement $u$ remains the only independent dynamic field, while the independent rotation $\theta$ must determined at each instant from its equilibrium equation. Equation \eqref{eq.Izero_Equation_Motion_1D} can be reduced to a single Timoshenko like PDE given by the following expression 
\begin{align}
	\left[\partial_t ^2 - c_{x2}^2 \partial_x ^2  + c_{x4}^4 \partial_x ^4  - c_{xt}^2 \partial_t ^2\partial_x ^2\right]  u = 0 ,
	\label{eq.General_EQ_zeroI}
\end{align}
If we assume an isotropic medium, we obtain the following terms 
\begin{align}
	c_{x2}^2 = \frac{\mu}{\rho}, \quad c_{x4}^4 = \left(\frac{\mu + \mu_c}{\rho}\right) \left(\frac{\mu L_c^2}{4\mu_c}\right), \quad c_{xt}^2=  \frac{\mu L_c^2}{4\mu_c}  .
	\label{eq.Micropolar_Single_Eq_Params2}
\end{align}
The coefficient $c_{x2}$ corresponds to the classical low-frequency shear-wave velocity. The parameter $c_{xt}^2$ has dimensions of length squared and controls the mixed space--time derivative, the parameter $c_{x4}^4/c_{xt}^2$ is limit of phase (and group) velocity at high frequencies.

\subsubsection{Dispersion analysis}

If we assume plane wave propagation (eq. \eqref{eq.Plane_Waves1}) we can obtain an expression for the phase velocity $(\omega/k)$ of the micropolar model with zero micro-inertia  given by
\begin{align}
	v_s^2 =
	\frac{
		c_{x2}^2
		-
		c_{xt}^2\omega^2
		+
		\sqrt{
			\left(
			c_{x2}^2
			-
			c_{xt}^2\omega^2
			\right)^2
			+
			4c_{x4}^4\omega^2
		}
	}{2},
	\label{eq.General_Phase_vel_zeroI}
\end{align}
which we an observe that it is the same phase velocity as the higher-order gradient model (see eq. \eqref{eq.Gradient_Phase_Velocity_Macro}). We can identify the two limiting cases
\begin{align}
	\begin{aligned}
		\text{macroscale} \to \lim_{\omega\to0}v_{s}^2 &= \frac{\mu}{\rho}=v_{\text{macro}}^2,\\
		\text{microscale} \to \lim_{\omega\to\infty}v_{s}^2 &= \frac{\mu + \mu_c}{\rho}=v_{\text{micro}}^2.
	\end{aligned}
	\label{eq.Phase_Velocity_New_Cosserat_limit}
\end{align}

We can next write a relationship between macroscopic and microscopic phase velocities, by combining eqs. \eqref{eq.Phase_Velocity_New_Cosserat_limit}, which results in the following expression
\begin{align}
	v_{\text{micro}}^2 = v_{\text{macro}}^2 + v_{\rm cosserat}^2 , \quad \text{with} \quad v_{\rm cosserat}^2 = \frac{ \mu_c}{\rho} .
	\label{eq.micropolar_scaling}
\end{align}

Equation \eqref{eq.micropolar_scaling} provides a mechanical relationship between micro and macro scales. Note we can write the difference of shear velocity between micro and macro scales as follows
\begin{align}
	v_{\text{micro}}^2-v_{\text{macro}}^2
	=
	v_{\rm cosserat}^2.
	\label{eq.micropolar_scaling2}
\end{align}

We can observe that the term $v_{\rm cosserat}=\sqrt{\mu_c/\rho}$, simply represents the increase in the
squared shear-wave velocity associated with the Cosserat couple modulus. Strictly speaking, it determines the difference between the microscopic and macroscopic limiting shear velocities. A larger value of $v_{\rm cosserat}$ results in a larger separation between the high-frequency ($v_{\rm micro}$) and low-frequency ($v_{\rm macro}$) limits, whereas $v_{\rm cosserat}\sim 0$ implies that both limits coincide and no dispersion is present.  

\subsection{Limitations of micropolar models}

Unlike the higher-order gradient model previously presented, the micropolar model (and all its variants) show a structural limitation, since eq. \eqref{eq.micropolar_scaling2} predicts that always $v_{\text{micro}}\geq v_{\text{macro}}$. Unlike the micropolar model, the higher-gradient model predicts that depending on a parameter choice of lenght scales, we can model: $v_{\text{micro}}\leq v_{\text{macro}}$ or $v_{\text{micro}}\geq v_{\text{macro}}$ (see eq. \eqref{eq.Micro_Macro_High_Gradient_relations}). This makes the higher-order gradient model much more general and useful for our purposes. 

\subsection{Challenges in full waveform modeling (and inversion)}

Observing eq. \eqref{eq.Gradient_Wave_Equation_1D}, the principal numerical difficulty arises from the mixed space--time derivative
\begin{align}
	\rho L_t^2 \partial_t^2 \partial_x^2 u .
\end{align}

Unlike the conventional elastic wave equation, here the acceleration term depends on the Laplacian. As a consequence, the unknown quantity at each time step is not simply $\partial_t^2u$, but rather the operator
\begin{align}
	\left(I-L_t^2\nabla^2\right)\partial_t^2u.
\end{align}

This increases the computational complexity enormously, since standard explicit finite-difference time-marching schemes cannot be directly applied \citep{igel2017computational}. Instead, at every time step one must solve an elliptic problem of the form
\begin{align}
	\left(I-L_t^2\nabla^2\right)\partial_t^2u=f,
\end{align}
where  $f$ contains the spatial elastic terms. This additional inversion considerably increases the computational cost, since an elliptic system must be solved at every time step.

A possible alternative can be evaluating the operator in the Fourier domain \citep{igel2017computational}. Under a pseudo-spectral discretization, the Laplacian becomes diagonal in wavenumber space. The acceleration can therefore be recovered through a simple division in Fourier space followed by an inverse transform. This approach avoids the repeated solution of large sparse linear systems and provides an efficient implementation of the dynamically consistent higher-gradient model for large-scale wave propagation problems. 

From a computational perspective, the development of efficient numerical strategies for large-scale full waveform modeling and inversion constitutes an important topic for future work.

\subsection{Implications for agroseismology}

Agroseismology is an emerging scientific field \citep{shi2026agroseismology,tsekhmistrenko2025between,muir2024quantifying}, whose long-term objective is to better understand the interaction between agricultural practices and soil health, in order to improve models of agricultural ecosystems and climate interactions. This will help to better understand how different farming practices affect soil health, detect harmful soil compaction early, to monitor crop-growing conditions (without digging or extensive sampling) and ultimately to improve agriculture and produce sustainable land management.

One of the main difficulties is that near-surface soils are highly heterogeneous and porous. Their mechanical response is influenced by processes occurring at several complex spatial scales. Classical homogenization approaches in seismology \citep[e.g.][]{capdeville20102,capdeville2013residual,capdeville2015fast,capdeville2018elastic,capdeville2007second,capdeville20101,hedjazian2021multiscale} attempt to eliminate the micro-scale by obtaining an effective elastic medium valid in the long-wavelength regime.

\begin{figure}
	\begin{center}
		\includegraphics[width=1\textwidth]{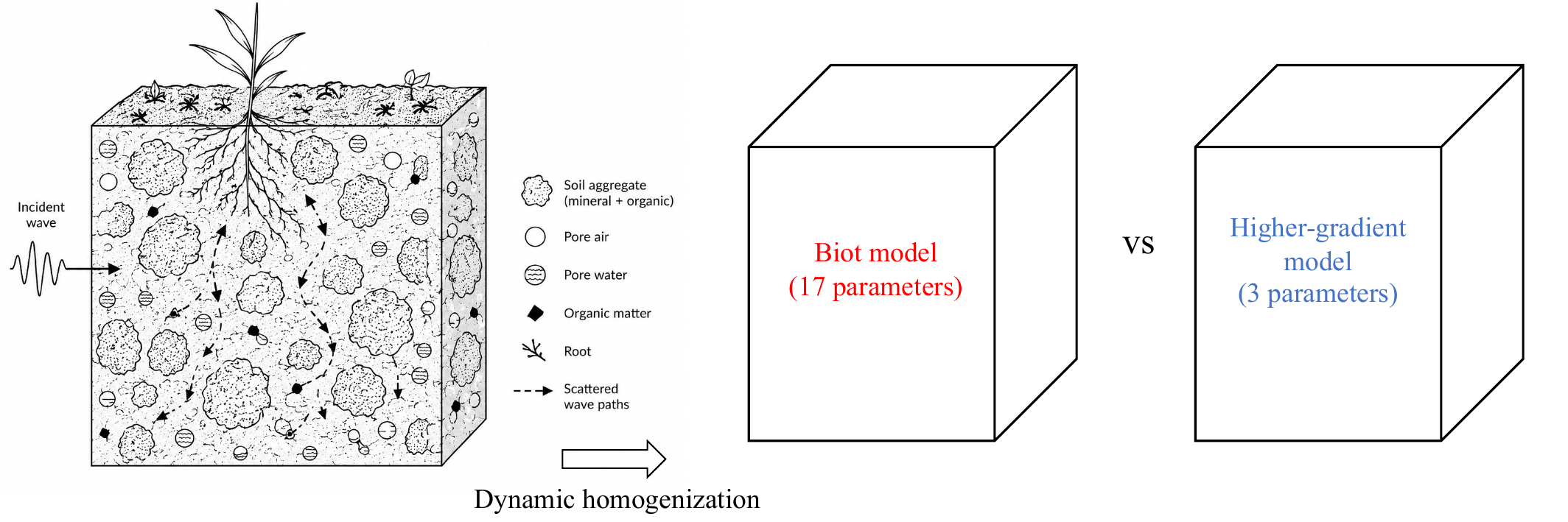}
		\caption{Dynamic homogenization of a poroelastic medium.}
		\label{fig.DAS}
	\end{center}
\end{figure}

We here attempt not to eliminate the microstructure but to reproduce its dynamics with a reduced medium of three parameters. We have shown that we are able to capture the experimentally observed shear-wave dispersion. From this perspective, the presented higher-order gradient model may provide an alternative framework for seismic inversion in porous soils. Instead of attempting to determine the large number of parameters required by modified Biot-type models, one may estimate a small set of effective micropolar parameters that capture the full dispersive behavior. This suggests a possible way forward to perform more realistic inversions for agroseismology (Fig.~\ref{fig.DAS}).

\section{Conclusions}

We have proposed a dynamically consistent higher-gradient continuum model for reproducing frequency-dependent shear-wave propagation in
heterogeneous porous media. The formulation introduces two independent characteristic lengths: an elastic characteristic length $L_c$ and an inertial characteristic length $L_t$.

The proposed model was validated against laboratory measurements of glycerin-saturated tight sandstone reported by \cite{yin2017pressure} over a broad range of effective pressures. Having only three effective material parameters, the model accurately reproduces the observed frequency-dependent shear-wave velocities across the complete experimental range. It successfully captures three distinct dispersion regimes: increasing phase velocity with frequency at low effective pressures, an almost dispersion-free response at intermediate pressures, and a weak decrease of phase velocity at the highest pressures. This behavior cannot generally be represented by Biot's models, which often require a large number of constitutive parameters describing pore geometry and fluid-flow mechanisms

These parameters should be interpreted as effective continuum quantities that summarize the collective influence of the underlying fluid--solid interactions, without requiring an explicit description of the pore-scale flow processes. Overall, the proposed dynamically consistent higher-gradient theory provides a simple, physically interpretable framework for modeling dispersive seismic wave propagation in porous rocks. 

Our results suggest that generalized continuum theories can offer an attractive reduced-order alternative to classical poroelastic models for applications in seismic exploration, reservoir characterization and agroseismology.

\section{Acknowledgments}

R.A. acknowledges constructive conversations with Fabian Bonilla. 

\section{Data availability}

Data used in this work can be downloaded from the supporting information of \cite{yin2017pressure}. No new experimental data were generated in this study. All equations required to reproduce the theoretical results are provided in the manuscript.

\section{Funding}

C.N. acknowledges financial support from the Institut de Physique du Globe de Paris (IPGP).

\section{Conflict of interest}

The authors have no conflicts of interest.

\footnotesize
\bibliographystyle{apalike}
\bibliography{Biblio}

\end{document}